\documentclass{article}
\usepackage{graphicx}
\def\be{\begin{equation}}
\def\ee{\end{equation}}
\def\bea{\begin{eqnarray}}
\def\eea{\end{eqnarray}}
\def\sh{\sinh}

\date{January 26, 2004}
\begin{document}
  \title{
    {\bf Buda-Lund hydro model and\\
      the elliptic flow at RHIC}
  }
  \author{M. Csan\'ad$^1$, T. Cs\"org\H{o}$^2$ and B. L\"orstad$^3$\\[1ex]
    {\small $^1$ Dept. Atomic Phys., ELTE, H-1117 Budapest XI, P\'azm\'any P. 1/A, Hungary} \\
    {\small $^2$ MTA KFKI RMKI, H - 1525 Budapest 114, P.O.Box 49, Hungary}\\
  {\small $^3$ Department of Physics, University of Lund, S-22362 Lund, Sweden}
  }
  \maketitle
  \begin{abstract}
    The ellipsoidally symmetric Buda-Lund hydrodynamic model
    describes naturally the transverse momentum and the
    pseudorapidity dependence of the elliptic flow 
    in Au+Au collisions at $\sqrt{s_{NN}} = 130$ and $200$ GeV.
    The result confirms the indication of quark deconfinement 
    in Au+Au collisions at RHIC, obtained from
    Buda-Lund hydro model fits to  combined 
    spectra and HBT radii  of BRAHMS, PHOBOS, PHENIX and STAR.
  \end{abstract}

\medskip
 
{\bf\it Introduction}.
PHENIX, PHOBOS and STAR experiments at RHIC produced a
wealth of information on the asymmetry of the particle spectra
with respect to the reaction 
plane~\cite{PHENIX-v2-id,PHENIX-v2-cent,PHOBOS-v2, PHOBOS-v2-qm02,STAR-v2-PRC,STAR-v2-PRL},
characterized by the second harmonic moment of the 
transverse momentum distribution, denoted by $v_2$. It is measured
as a function of the transverse mass and particle type 
at mid-rapidity as well as a function of the pseudo-rapidity 
$\eta = 0.5 \log(\frac{|p| + p_z}{|p| - p_z})$.

The PHOBOS collaboration found~\cite{PHOBOS-v2},
that $v_2(\eta)$ is a strongly decreasing function
of $|\eta|$, which implies
that the concept of boost-invariance, suggested by
Bjorken in ref.~\cite{Bjorken}, cannot be applied to characterize
the hadronic final state of Au+Au collisions at RHIC.

We summarize here a successful attempt to describe
the pseudo-rapidity dependence of the elliptic flow $v_2(\eta)$ at RHIC,
for more details see ref.~\cite{mate-ell1}. Our tool is the Buda-Lund
hydrodynamic model~\cite{3d,3d-qm95},
which we extended in ref.~\cite{mate-ell1} from axial to ellipsoidal symmetry.

{\bf\it Buda-Lund hydro for ellipsoidal expansions}.  
Based on the success of the Buda-Lund hydro model to describe
$Au+Au$ collisions at RHIC~\cite{bl-rhic,ster-ismd03}, $Pb+Pb$ collisions
at CERN SPS~\cite{bl-sps} and
$h+p$ reactions at CERN SPS~\cite{bl-na22,cs-rev},
we describe the emission function in the core-halo picture, and
assume that the core evolves in a  hydrodynamical manner:
\begin{equation}
  S_c(x,p) d^4 x = \frac{g}{(2 \pi)^3}
  \frac{ p^\mu d^4\Sigma_\mu(x)}{B(x,p) +s_q},
\end{equation}
where $g$ is the degeneracy factor ($g = 1$ for identified 
pseudoscalar mesons, $g = 2$ for identified spin=1/2 baryons),
and $p^\mu d^4 \Sigma_\mu(x)$ is a generalized Cooper-Frye term,
describing the flux of particles through
a distribution of layers of freeze-out hypersurfaces,
$B(x,p)$ is the (inverse) Boltzmann phase-space distribution,
and the term $s_q$ is determined by quantum statistics,
$s_q = 0$, $-1$, and $+1$  for Boltzmann, Bose-Einstein and Fermi-Dirac
distributions, respectively.

For a hydrodynamically expanding system, the (inverse) Boltzmann
phase-space distribution is
\begin{equation}B(x,p)=
  \exp\left( \frac{ p^\nu u_\nu(x)}{T(x)} -\frac{\mu(x)}{T(x)} \right).
\end{equation}
We will utilize some ansatz for the shape of the flow four-velocity,
$u_\nu(x)$, chemical potential, $\mu(x)$, and temperature, $T(x)$
distributions. Their form is determined with
the help of recently found exact solutions of hydrodynamics,
both in the relativistic~\cite{relsol-cyl,relsol-ell}
and in the non-relativistic cases~\cite{nr-sol,nr-ell,nr-inf}.

The generalized Cooper-Frye prefactor is
determined from the assumption that the freeze-out happens, with
probability $H(\tau) d\tau$, at
a hypersurface characterized by $\tau=const$ and
that the proper-time measures the time elapsed in a fluid element
that moves together with the fluid, $d\tau = u^\mu(x) dx_\mu$.
We parameterize this hypersurface with the coordinates $(r_x,r_y,r_z)$
and find that $d^3 \Sigma^\mu(x|\tau) = u^\mu(x) d^3 x/ u^0(x)$.
Using $\partial_t \tau|_r = u^0(x)$ we find that in this case
the generalized Cooper-Frye prefactor is
\begin{equation}
  p^\mu d^4 \Sigma_\mu(x) =p^\mu u_\mu(x) H(\tau) d^4 x,
\end{equation}
This finding generalizes a  
result of ref.~\cite{cracow} from the case of a spherically
symmetric Hubble flow to anisotropic, direction dependent Hubble
flow distributions.

From the analysis of CERN SPS and RHIC data~\cite{bl-sps,bl-rhic,ster-ismd03},
we find that the proper-time distribution in heavy ion collisions
is rather narrow, and $H(\tau)$ can be well
approximated with a Gaussian representation of the Dirac-delta
distribution,
\begin{equation}
  H(\tau) = \frac{1}{(2 \pi \Delta\tau^2)^{1/2}}
  \exp\left(-\frac{(\tau - \tau_0)^2 }{ 2 \Delta \tau^2}\right),
\end{equation}
with $\Delta \tau \ll \tau_0$.

We specify a fully scale invariant,
relativistic form, which reproduces
known non-relativistic hydrodynamic solutions too,
in the limit when the expansion is non-relativistic.
Both in the relativistic and the non-relativistic cases, 
the ellipsoidally symmetric, self-similarly expanding 
hydrodynamical solutions can be formulated in
a simple manner, using a scaling variable $s$ and a 
corresponding four-velocity distribution $u^\mu$, that satisfy
\begin{equation}
  u^\mu \partial_\mu s = 0, \label{e:scalingvar}
\end{equation}
which means that $s$ is a good scaling variable
if its co-moving derivative vanishes~\cite{relsol-ell,relsol-cyl}.

It is convenient to introduce the dimensionless,
generalized space-time rapidity variables $(\eta_x,\eta_y,\eta_z)$,
defined by the identification of 
\be 
	\sh \eta_x = r_x \frac{\dot X}{X},
\ee
similar equations hold for $y$ and $z$.  
The characteristic sizes (for example, the lengths of
the major axis of the expanding ellipsoid)
are $(X,Y,Z)$  that depend on proper-time $\tau$ and
their derivatives with respect to proper-time are denoted by
$(\dot X,  \dot Y, \dot Z)$.
Eq.~(\ref{e:scalingvar}) is satisfied by the choice of
\begin{eqnarray}
  s & = & \frac{\cosh \eta_x -1}{\dot X^2_f}
  +\frac{\cosh \eta_y -1}{\dot Y^2_f}
  +\frac{\cosh \eta_z -1}{\dot Z_f^2}, \\
  u^\mu & = & (\gamma, \,  \sinh \eta_x, \, \sinh \eta_y, \, \sinh\eta_z),
\end{eqnarray}
and from here on
$(\dot X_f, \dot Y_f, \dot Z_f)
= (\dot X(\tau_0), \dot Y(\tau_0), \dot Z(\tau_0) )
= (\dot X_1, \dot X_2,\dot X_3)$,
assuming that the rate of expansion is constant in the narrow proper-time
interval of the freeze-out process. The above form has the desired
non-relativistic limit,
\begin{equation}
  s \rightarrow
  \frac{r_x^2}{2 X_f^2} + \frac{r_y^2}{2 Y_f^2} +\frac{r_z^2}{2 Z_f^2},
\end{equation}
where again
$(X_f, Y_f, Z_f) = (X(\tau_0), Y(\tau_0), Z(\tau_0))=(X_1,X_2,X_3)$.
From now on, we drop subscript $_f$.
The normalization condition of $u^\mu(x) u_\mu(x) = 1$
yields the value of $\gamma$.
For the fugacity distribution we assume a shape, that leads to
Gaussian profile in the non-relativistic limit,
\begin{equation}
  \frac{\mu(x)}{T(x)} = \frac{\mu_0}{T_0} - s,
\end{equation}
corresponding to the solution discussed
in refs.~\cite{nr-sol,nr-ell,csorgo-ellobs}.
We assume that the temperature may depend on the position as
well as on proper-time.
We characterize the inverse temperature distribution
similarly to the shape used in the axially symmetric model of
refs.~\cite{3d,3d-qm95}, and discussed in the exact hydro
solutions of refs.~\cite{nr-sol,nr-ell},
\begin{equation}\frac{1}{T(x)}= \frac{1}{T_0} \left( 1 +
  \frac{T_0 - T_s}{T_s} \:s\right) \left( 1 + \frac{T_0 - T_e}{T_e}  \,
  \frac{(\tau -\tau_0)^2}{2 \Delta\tau^2}\right)
\end{equation}
where $T_0$, $T_s$ and $T_e$ are the temperatures of the center,
and the surface at the mean freeze-out time $\tau_0$,
while $T_e$ corresponds to the temperature of the
center after most of the particle emission is over (cooling 
due to evaporation and expansion). Sudden emission implies 
$T_e = T_0$ and $\Delta\tau \rightarrow 0$.

The observables can be calculated analytically from the Buda-Lund
hydro model, using a saddle-point approximation in the integration. 
This approximation  is exact both in the
Gaussian and the non-relativistic limit,
and if $p^\nu u_\nu / T \gg 1$ at the point of
maximal emittivity.

The results are summarized in Figs. 1 and 2. We find that
a small asymmetry between the two transverse Hubble constants 
gives a natural explanation of the transverse momentum dependence of $v_2$.
The parameters are taken from Buda-Lund hydro model fits in
refs.~\cite{bl-rhic,ster-ismd03}, where the axially symmetric version of the
model was utilized, although the mean transverse flow is reduced due to
the less central nature of collisions studied in this study, which is 
a summary of ref.~\cite{mate-ell1}. Note that we did not yet fine-tune
the ellipsoidal Buda-Lund hydro model to describe these $v_2$ data, 
instead we searched the parameter set by hand. 
So at the moment we are not yet ready to
report the best fit parameters and the error bars on the extracted
parameter values.  However, as indicated by Figs. 1 and 2,
even this fit by eye method is successful in reproducing the data 
on elliptic flow at RHIC.

First we tuned the model to describe the $p_t$ dependent elliptic
flow of identified particles at midrapidity, as shown in Fig. 1. 
Then we calculated the value of the
transverse momentum integrated $v_2(\eta=0)$ and found~\cite{mate-ell1},
that this value is below the published PHOBOS data point at mid-rapidity.
We attribute the difference of 0.02 to a non-flow
contribution~\cite{borghini,STAR-v2-PRC}. 
The PHOBOS collaboration pointed out the possible existence of
such a non-flow contribution in their data  in ref.~\cite{PHOBOS-v2},
as they did not utilize the fourth order cumulant measure of 
$v_2$. 

We note that our presently best choice of  parameter set correspond to a
high, $T_0 > T_c = 170 $ MeV central temperature, with a cold surface
temperature of $T_s \approx 105$ MeV, see Figs. 1 and  2.

{\bf\it Summary and conclusions}.  
We have generalized the Buda-Lund hydro model to the case of
ellipsoidally symmetric expanding fireballs.
We kept the parameters as determined from fits to the single particle
spectra and the two-particle Bose-Einstein correlation functions
(HBT radii)~\cite{bl-rhic,ster-ismd03},
and interpreted them as angular averages over the 
direction of the reaction plane.
Then we observed that a small splitting between the expansion rates
parallel and transverse to the direction of the impact parameter,
as well as a small   tilt of the particle emitting source
is sufficient to describe simultaneously the
transverse momentum dependence of the collective flow
of identified particles~\cite{PHENIX-v2-id} as well as
the pseudorapidity dependence of the collective flow
~\cite{PHOBOS-v2,PHOBOS-v2-qm02} at RHIC.

The results confirm the indication for quark deconfinement at
RHIC found in refs.~\cite{bl-rhic,ster-ismd03},
based on the observation, that some of the particles are emitted
from a region with higher than the critical temperature,
$T > T_c = 170$ MeV.  The size of this
volume is about 750 fm$^3$, corresponding to 1/8-th of the 
total volume measured on the $\tau=\tau_0$ main freeze-out 
hypersurface~\cite{mate-ell1}.
At the same time, this analysis indicates that the
surface temperature is rather cold, $T_s \approx 105$ MeV,
so approximately 7/8 of the particles are emitted from a rather cold
hadron gas. So the picture is similar to a fireball, 
which is heated from inside.

{\bf\it Acknowledgments}.
T. Cs. and M. Cs. would like to the Organizers for their kind hospitality
and for their creating an
inspiring and fruitful meeting. 
The support of the following grants are gratefully acknowledged:
OTKA T034269, T038406, the OTKA-MTA-NSF grant INT0089462, 
the NATO PST.CLG 980086 grant and 
the exchange program of the Hungarian and Polish Academy of Sciences.

\begin{figure}
  \begin{center}
    \includegraphics*[scale = 0.7]{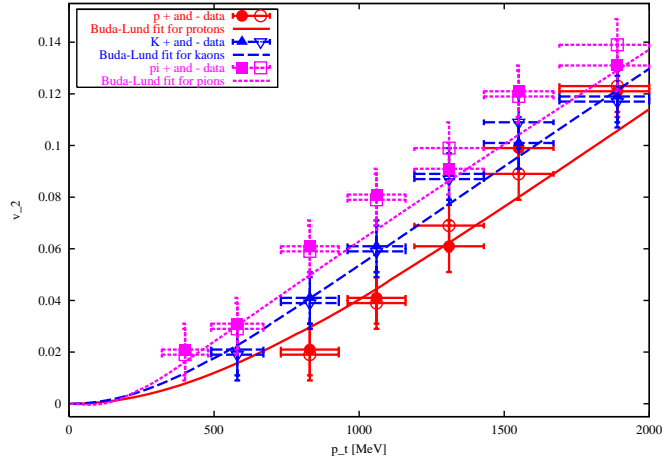}
  \end{center}
  \caption{Buda-Lund fit to the $v_2(p_t)$ data}\label{v2pt}
  \small{Here we see the fit to the PHENIX $v_2(p_t)$ data of
    identified particles~\cite{PHENIX-v2-id}. The parameter set is:
    $T_0=210\textrm{ MeV}$, $\dot X=0.57$, $\dot Y=0.45$, $\dot
    Z=2.4$, $T_s=105\textrm{ MeV}$, $\tau_0 = 7$ fm/c, $\vartheta=0.09$, $X_f=8.6$ fm,
    $Y_f=10.5$ fm, $Z_f=17.5$ fm, $\mu_{0,\pi}=70\textrm{ MeV}$,
    $\mu_{0,K}=210 \textrm{ MeV}$ and $\mu_{0,p}=315 \textrm{ MeV}$,
    and the masses are taken as their physical value.}
\end{figure}

\begin{figure}
  \begin{center}
    \includegraphics*[scale = 0.7]{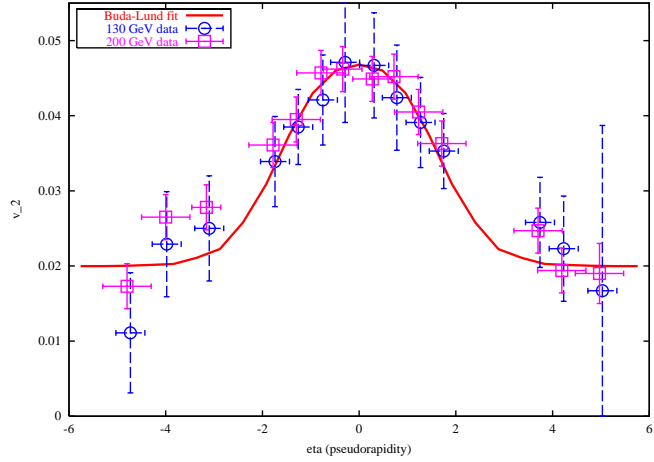}
  \end{center}
  \caption{Buda-Lund fit to the $v_2(\eta)$ data}\label{v2y}
  \small{This image shows the fit to the $130 \textrm{ GeV}$
    Au+Au and $200 \textrm{ GeV}$ Au+Au  $v_2(\eta)$
    data of PHOBOS~\cite{PHOBOS-v2,PHOBOS-v2-qm02},
    with the ellipsoidal generalization of the Buda-Lund hydro model.
    Here we used the same parameter set as at fig. \ref{v2pt}, 
    with pion mass and chemical potential, and a constant non-flow parameter of 0.02.}
\end{figure}


\end{document}